# Neural Sampling Machine with Stochastic Synapse allows Brain-like Learning and Inference

Sourav Dutta[1*], Georgios Detorakis[2], Abhishek Khanna[1], Benjamin Grisafe[1], Emre Neftci[2] and Suman Datta[1]

[1]Department of Electrical Engineering, University of Notre Dame, Notre Dame, IN 46556, USA
[2]Department of Cognitive Sciences, University of California Irvine, Irvine, CA, 92697, USA
*Corresponding author: sdutta4@nd.edu

**Many real-world mission-critical applications require continual online learning from noisy data and real-time decision making with a defined confidence level. Probabilistic models and stochastic neural networks can explicitly handle the uncertainty in data and allowing adaptive learning on the fly, but their implementation in a low-power substrate remains a challenge. In this work, we introduce a novel hardware fabric that can implement a new class of stochastic neural network called Neural Sampling Machine (NSM) that exploits stochasticity in its synaptic connections for approximate Bayesian inference. Harnessing the inherent non-linearities and stochasticity occurring at the atomic level in emerging materials and devices allows us to capture the synaptic stochasticity occurring at the molecular level in biological synapses. We experimentally demonstrate an *in silico* hybrid stochastic synapse by pairing a ferroelectric field-effect transistor (FeFET)-based analog weight cell with a two-terminal stochastic selector element. Such a stochastic synapse can be integrated within the well-established crossbar array architecture for compute-in-memory (CIM). We experimentally show that the inherent stochastic switching of the selector element between the insulator and metallic state introduces a multiplicative stochastic noise within the synapses of NSM that samples the conductance states of the FeFET, both during learning and inference. Using experimentally calibrated models, we perform network-level simulations to highlight the salient automatic weight normalization feature introduced by the stochastic synapses of the NSM that paves the way for continual online learning without any offline Batch Normalization. We also showcase the Bayesian inferencing capability introduced by the stochastic synapse during inference mode, thus accounting for uncertainty in data. We report high accuracy of 98.25% on standard image classification task as well as the estimation of data uncertainty in original vs. rotated samples. Building such a stochastic NSM hardware will allow using inspiration from neuroscience to design a ML architecture that can learn and report uncertainty.**

Harnessing the intricate dynamics at the microscopic level in emerging materials and devices have unraveled new possibilities for brain-inspired computing such as building analog multi-bit synapses [1]–[10] and bio-inspired neuronal circuits [10]–[12]. Such emerging materials and devices also exhibit inherent stochasticity at the atomic level which is often categorized as a nuisance for information processing. In contrast, variability is a prominent feature exhibited by biological neural networks at the molecular level are believed to contribute to the computational strategies of the brain [13]. Such variability has been reported in the recordings of biological neurons or as unreliability associated with the synaptic connections. Typically, a presynaptic neuronal spike causes the release of neurotransmitters at the synaptic release site as illustrated in Fig. 1(a). Borst *et. al.* [14] reported that the synaptic vesicle release in the brain can be extremely unreliable. The transmission rate can be as high as 50% and as low as 10% measured *in vivo* at a given synapse. Synaptic noise has the distinguishing feature of being *multiplicative* which plays a key role in learning and probabilistic inference dynamics. In this work, we propose a novel stochastic synapse that harnesses the inherent variability present in emerging devices and mimic the dynamics of a noisy biological synapses. This allows us to realize a novel neuromorphic hardware fabric that can support a recently proposed class of stochastic neural network called the Neural Sampling Machine (NSM) [15].

While the functional role of this multiplicative stochasticity in the brain is still under debate, the biologically inspired stochasticity can be exploited in certain machine learning algorithms. In particular, NSMs build on the idea of introducing stochasticity at various levels in a neural network to allow – (a) escaping local minima during learning and inference [16], (b) regularization in neural networks [17], [18], (c) approximate Bayesian inference with Monte Carlo sampling [19], [20] and (d) energy efficient communication and computation [21], [22]. NSM draws inspiration from regularization techniques such as Dropout [17] or DropConnect [18] that randomly drop a subset of neural activation or weights in the neural network during the forward pass of training. Contrary to DropConnect where stochasticity is switched off during inference, the synaptic stochasticity is always present in an NSM. This *always-on* stochasticity confers probabilistic inference capabilities to the network [20] and is consistent with the idea of continual learning and lifelong learning machines while improving energy efficiency [21], [22]. Neural networks equipped with *always-on* stochasticity have been shown to match or surpass the performance of contemporary machine learning algorithms. Together with multiplicative noise incorporated in stochastic synapses, this new class of NSM provides an important pathway towards realizing probabilistic inference [23] and active learning [24], [25].

In this work, we propose a hardware implementation of NSM using hybrid stochastic synapses consisting of an embedded non-volatile memory (eNVM) in series with a two-terminal stochastic selector element. We experimentally demonstrate *in silico* such a hybrid stochastic synapse by pairing a doped $HfO_2$ FeFET-based analog weight cell with a two-terminal $Ag/HfO_2$ stochastic selector. Such hybrid synapses can be integrated within the prevailing crossbar array architecture for CIM that provides a promising energy-efficiency pathway for building neuromorphic hardware by reducing data-movement[26]. We exploit the inherent stochastic switching of the selector element between the insulator and the metallic state to perform Bernoulli sampling of the conductance states of the FeFET both during learning and inference. A remarkable side-effect of the multiplicative noise dynamics is a self-normalizing effect that performs automatic weight normalization and prevention of internal covariate shift in an online fashion. Furthermore, the

*always*-on stochasticity of the NSM during the inference mode allows performing Bayesian inferencing.

**Theoretical Model of NSM**

Neural Sampling Machines (NSM) are stochastic neural networks that exploit neuronal and/or synaptic noise to perform learning and inference [15]. A schematic illustration is shown in Fig. 1(b) comprising synaptic stochasticity that injects a multiplicative Bernoulli or "*blank-out*" noise in the model. Such a noise can be incorporated in the model as a continuous DropConnect [18] mask on the synaptic weights such that a subset of the weights is continuously forced to be zero as shown in Fig. 1(b). Next, we lay down a theoretical description of the NSM.

We use binary threshold neurons with the following activation function

$$z_i = sgn(u_i) = \begin{cases} -1, & if \ u_i < 0 \\ 1, & if \ u_i \geq 0 \end{cases} \quad (1)$$

where $u_i$ is the pre-activation of neuron $i$ and is given by:

$$u_i = \sum_{j=1}^{N}(\xi_{ij}+a_i)w_{ij}z_j + b_i \quad (2)$$

where $w_{ij}$ represents the weight of the synaptic connection between neurons $i$ and $j$ and $\xi_{ij}$ is the multiplicative Bernoulli noise modeled using an independent and identically distributed (*iid*) random variable with parameter $p$ such that $\xi_{ij} \sim Bernoulli(p) \in [0,1]$. $b_i$ is a bias term applied per neuron $i$. An additional term $a_i$ is added per neuron $i$ to counter the scaling factor issue due to multiplicative noise [27]. It can be further shown that for such binary threshold neurons, the probability of a neuron firing is given by

$$P(z_i = 1|\mathbf{z}) = \frac{1}{2}\left[1 + \text{erf}\left(\frac{\mathbb{E}(u_i|\mathbf{z})}{\sqrt{2Var(u_i|\mathbf{z})}}\right)\right] \quad (3)$$

where $\mathbb{E}(u_i)$ and $Var(u_i)$ are the expectation and variance of $u_i$. For Bernoulli type noise, the probability of neuron firing becomes [27]

$$P(z_i = 1|\mathbf{z}) = \frac{1}{2}\left[1 + \text{erf}\left(\frac{(p+a_i)\sum_j w_{ij}z_j}{\sqrt{2p(1-p)\sum_j w_{ij}^2}}\right)\right] = \frac{1}{2}\left[1 + \text{erf}\left(\frac{(p+a_i)\sum_j w_{ij}z_j}{\sqrt{2p(1-p)}\|\mathbf{w}_i\|}\right)\right]$$

$$= \frac{1}{2}\left[1 + \text{erf}\left(\beta_i \frac{\sum_j w_{ij}z_j}{\|\mathbf{w}_i\|}\right)\right] = \frac{1}{2}[1 + \text{erf}(\mathbf{v}_i \cdot \mathbf{z})] \quad (4)$$

with $\beta = \frac{p+a_i}{\sqrt{2p(1-p)}}$ capturing the noise in the model and $v_i = \beta_i \frac{\mathbf{w}_i}{\|\mathbf{w}_i\|}$. Here, $\|\cdot\|$ denotes the L2 norm of the weights of neuron $i$. Note that the notion behind weight normalization is to reparameterize the weight vector using $v_i = \beta_i \frac{\mathbf{w}_i}{\|\mathbf{w}_i\|}$ [28] which is exactly the same as that obtained in NSM due to the inherent stochastic noise in the synapses. Thus, NSM inherently introduces the salient self-normalizing feature and performs weight normalization in the same

sense as [28]. One important feature of the NSM is that since this weight normalization is an inherent feature of the model, NSM offers the features equivalent to batch normalization in an online fashion. Additionally, by decoupling the magnitude and the direction of the weight vector, a potential speedup in convergence is obtained [27].

**Implementing NSM using Emerging Devices Operating in Stochastic Switching Regime**

Recent years have seen extensive research on building dedicated hardware for accelerating DNNs using CIM approach. The core computing kernel consists of a crossbar array with perpendicular rows and columns with eNVMs placed at each cross-point as shown in Fig. 1(c). The weights in the DNN are mapped to the conductance states of the eNVM. The crossbar array performs row-wise weight update and column-wise summation operations in a parallel fashion as follows: the input (or read) voltages $\boldsymbol{V_{in}}$ from the input neuron layer are applied to all the rows and are multiplied by the conductance of the eNVM at each cross-point $G$ to create a weighted sum current in each column $\boldsymbol{I_{out}} = \sum G \boldsymbol{V_{in}}$. The output neuron layer placed at the end of the column converts these analog currents into digital neuronal outputs.

Implementing an NSM with the same existing hardware architecture requires selectively sampling or reading the synaptic weights $G_{ij}$ with some degree of uncertainty, based on random binary variables $\xi_{ij}$ generated for each of the synapse. We show that this can be easily realized by pairing the eNVM in series with a two-terminal stochastic selector element at each cross-point as shown illustratively in Fig. 1(c). We choose a selector device such that it operates as a switch, stochastically switching between an ON state (representing $\xi_{ij} = 1$) and an OFF state ($\xi_{ij} = 0$). The detailed description of such a selector is mentioned later. Fig. 1(d) shows a scenario where an input voltage $V_{in,3}$ is applied to the third row of the synaptic array while the conductance of the synapses are set to $\boldsymbol{G} = \{G_1, G_2, G_3, G_4, \dots, G_N\}$. Depending on the state of the selectors in the cross-points, an output weighted sum current $\boldsymbol{I_{out}} = \{0, G_2 V_{in,3}, 0, G_4 V_{in,3}, \dots, 0\}$ is generated. This is exactly the same as multiplying the weight sum of $w_{ij} z_j$ with a multiplicative noise $\xi_{ij}$ as described in Eqn. 2.

**Building Blocks for Stochastic Synapse: FeFET-based Analog Weight Cell**

The idea of voltage-dependent partial polarization switching in ferroelectric $Hf_xZr_{1-x}O_2$ can be leveraged to implement a non-volatile FeFET-based analog synapse. The FeFET-based synapse can be integrated into a pseudo-crossbar array that is suitable for row-wise weight update and column-wise summation [6], [10], [29]. Fig. 2(a) shows the schematic of a FeFET-based analog synapse without any additional stochastic selector element. The channel conductance $G$ of the FeFET can be modulated by applying write voltage pulses $\pm V_{write}$ to the gate of the FeFET. For reading out the conductance state, a small read voltage $V_{read}$ is applied to the gate terminal. With an input voltage $V_{in}$ applied to the drain of the FeFET, the output (drain) current becomes $I_{out} = G V_{in}$. Fig. 2(b) shows the experimentally measured conductance modulation in a 500 nm x 500 nm high-K metal gate FeFET fabricated at 28nm technology node [30]. For on-line learning on crossbar arrays, typically potentiation and depression pulse schemes with identical pulse amplitudes and widths are preferred. Nonetheless for a proof-of-concept, we used an amplitude modulation scheme where write voltage pulses $V_{write}$ of increasing amplitude from 2.8V to 4V and pulse widths of $1\mu s$ are applied to modulate the conductance of the FeFET. Applying progressively increasing negative pulses causes the FeFET to transition from the initial low resistance state (LRS) with lower threshold voltage ($V_T$) to high resistance state (HRS) as shown by the current-voltage characteristics in Fig. 2(b). Similarly, applying progressively increasing positive pulses causes a change in the conductance from HRS to LRS. Fig. 2(c) shows a continuous change in the conductance state of the FeFET upon applying multiple potentiation and depression pulses of varying amplitude and constant pulse width of $1\mu s$. The cycle-to-cycle variation in the measured conductance states observed in Fig. 2(c) arises due to the inherent stochastic switching dynamics of the individual ferroelectric domains [31]. Such inherent stochasticity also results in a device-to-device variation of the conductance states. To incorporate such variability, we measured the conductance modulation both for potentiation and depression across ten devices as shown in Fig. 2(d). We incorporate the model of FeFET-based analog weight cell in the NSM by fitting the conductance update scheme for both potentiation and depression with the closed-form expression $\Delta G = \alpha + \beta \left(1 - e^{-(|V_{write}|-V_0)/\gamma}\right)$ where $\alpha, \beta, \gamma$ and $V_0$ are the fitting parameters.

**Building Blocks for Stochastic Synapse: Ag/HfO$_2$ Stochastic Selector**

Next, we describe the characteristics of our stochastic selector device. Fig. 3(a) shows a schematic and a transmission electron microscopy (TEM) of a fabricated stack of [Ag/TiN/HfO$_2$/Pt] with 3nm TiN and 4nm HfO$_2$. A stochastic synapse is realized by augmenting this stochastic selector in series with the FeFET-based analog weight cell as shown in Fig. 2(b). The [Ag/TiN/HfO2/Pt] metal ion threshold switch device, from here on referred to as the Ag/ HfO$_2$ selector device, operates based on the principle of metal ion migration through a metal oxide medium similar to conducting bridge RAM (CBRAM). Starting from an initial OFF state, under an applied external bias, Ag atoms ionize and respond to the electric field migrating via interstitial hopping from top electrode to bottom electrode until a continuous filament of Ag+ atoms bridge the top and bottom electrodes. This is accompanied by several orders of magnitude change in conductivity as the device turns ON as shown by the measured current-voltage characteristics in Fig. 3(c). As the field is reduced, the inclination for Ag atoms to form clusters with other Ag atoms, rather than linear chains of atoms in contact with Pt allows for the spontaneous rupture of the atomic filament, turning OFF the device [32]. The role of TiN in the stack is to limit the initial migration of Ag during the electroforming sweep, such that device reliability is enhanced [33]. Interestingly, upon repeated measurement of the switching characteristic of the selector device, we see a considerable variation in the threshold voltage $V_T$ that triggers the spontaneous formation of the Ag$^+$ filament through HfO$_2$ insulating matrix and turning ON the device as shown in Fig. 3(d). Such stochastic switching can be exploited by applying the input voltage $V_{in}$ within the variation window of the $V_T$ as shown in Fig. 3(c). This would allow stochastic sampling of the conductance state of the FeFET in series. Figs. 3(e) and (f) show two examples of stochastically reading an LRS and an HRS of the FeFET through the stochastic selector. Overall, this validates the proposed idea of using such a hybrid structure as a truly stochastic synapse for implementing NSM on the hardware.

The stochasticity switching of the selector device is incorporated in the NSM by modeling it as an Onrstein-Uhlenbeck (OU) Process. The dynamics of the $V_T$ can be described as

$$dV_T = \theta(\mu - V_T)dt + \sigma dW \tag{5}$$

where W is the Wiener process, $\theta$ describes the magnitude of the mean-reverting force towards the mean $\mu$. $\sigma$ captures the diverting variance. The calibrated OU process shows excellent agreement with our experimental results as shown in Figs. 3(g)-(i) in terms of the cycle-to-cycle variation of $V_T$, overall distribution of $V_T$ and autocorrelation. Details of the OU calibration is included in the Methods section.

**Hardware NSM and Image Classification Task**

We test the performance of our hardware NSM incorporating FeFET-based analog weight cell and stochastic selector as the hybrid stochastic synapse on image classification task using the MNIST handwritten digit dataset as an example. Fig. 4(a) shows the network architecture consisting of an input layer with 784 neurons, three fully connected hidden layers with 300 neurons and a softmax output layer of 10 neurons for 10-way classification. For comparison, we chose three networks with the same architecture – (a) deterministic feedforward multilayer perceptron (MLP), (b) theoretical NSM model with full precession synaptic weights and a Bernoulli multiplicative noise for the stochastic synapses and (c) simulated hardware-NSM using the FeFET-based analog weight cell and the stochastic selector. The hardware NSM is trained using backpropagation and a softmax layer with cross-entropy loss and minibatch size of 100. While training of the hardware NSM, during the backward pass, the weight update is applied using the derivative of Eqn. (4) and the closed-form equation in Fig.2(d). During the forward pass for both learning and inference, the weights are stochastically accessed. This involves calculating the $V_T$ of each selector device in the cross-points in every iteration using the OU process described by Eqn. 5 and constructing a Boolean matrix $\Xi$ such that if $V_T \geq V_{T,mean}$, $\xi_{ij} = 1$, else $\xi_{ij} = 0$. Subsequently, we evaluate Eqns. 1-2.

The exact nature of the multiplicative noise injected by the stochastic selector is understood by comparing the measured switching probability with the theoretically predicted probability of switching for a Bernoulli process. Fig. 4(b) shows an exact match between the measured and theoretically predicted probability, highlighting that our stochastic selector device can inject Bernoulli multiplicative noise. Fig. 4(c) and (d) shows the performance of the hardware NSM in terms of the test accuracy and comparison with the theoretical NSM model and conventional MLP network. It is seen that the theoretical model outperforms the conventional MLP network as highlighted in [27]. The simulated hardware-NSM shows comparable test accuracy with the conventional MLP, the performance mainly limited by the dynamic range and non-idealities of the FeFET-based synaptic weight cell.

**Inherent Weight Normalization and Robustness to Weight Fluctuations**

As explained earlier, NSM allows decoupling the weight matrix as $v_i = \beta_i \frac{w_i}{\|w_i\|}$ which provides several advantages. Firstly, an inherent weight normalization can be effectively achieved without resorting to any batch normalization technique by performing gradient descent (calculating derivatives) with respect to the variables $\beta$ in addition to the weights $w$ as [27].

$$\frac{\partial \mathcal{L}}{\partial \beta_i} = \frac{\sum_j w_{ij} \partial_{v_{ij}} \mathcal{L}}{\|w_i\|} \tag{6}$$

$$\frac{\partial \mathcal{L}}{\partial w_{ij}} = \frac{\beta_i}{\|w_i\|} \frac{\partial \mathcal{L}}{\partial v_{ij}} - \frac{\beta_i}{\|w_i\|^2} w_i \frac{\partial \mathcal{L}}{\partial \beta_i} \tag{7}$$

This allows the distribution of the weights in the NSM to remain more stable than a conventional MLP without any additional weight regularization applied. Fig. 4(e) shows the evolution of the weights of the third layer during learning for three cases – (a) an MLP without any regularization, (b) MLP with additional regularization added and (c) hardware NSM. It is seen that the distribution of NSM weights is narrower and remains concentrated around its mean (low variance). On the other hand, the variance of the weight distribution is larger for the MLP network without weight regularization.

**Mitigation of Internal Covariate Shift**

The self-normalizing feature of the NSM also prevents the internal covariate shift caused by changes in the input distribution, similar to that achieved using Batch normalization. To highlight this, we next compare the 15th, 50th and 85th percentiles of the input distributions to the last hidden layer during training for all the three networks as shown in Fig. 4(f). The internal covariate shift is clearly visible in the conventional MLP without any normalization incorporated as the input distributions change significantly during the learning. In contrast, the evolution of the input distribution in the hardware NSM is remains stable, suggesting that NSMs prevents internal covariate shift through the self-normalizing effect that inherently performs weight normalization.

**Bayesian Inferencing and Capturing Uncertainty in Data**

Next, we showcase the ability of our simulated hardware-NSM to perform Bayesian inferencing and produce classification confidence. For this, we train our hardware NSM on the full MNIST dataset. During the inference mode, we evaluate the performance of the trained NSM on continuously rotated images of the digits 1 and 2 and shown in Fig. 5(a) and (f). For each of the rotated images, we perform 100 stochastic forward passes and record the softmax input (output of the last fully connected hidden layer in Fig. 4(a)) as well the softmax output. We highlight the response of 3 representative neurons - 1, 2 and 4 out of all the 10 neurons that show the highest activity. It is seen that when the softmax input of a particular neuron is larger than all the other neurons, the NSM will predict the class corresponding to that neuron. For example, in Fig. 5(b)-(d), for the first seven images, the softmax input for neuron 1 is largest. Consequently, the softmax output for neuron 1 remains close to 1 and the NSM predicts the images as belonging to class 1. However, as the images are rotated more, it is seen that even though the softmax output can be arbitrarily high for neuron 2 or 4 predicting that the image belongs to the class 2 or 4, respectively, the uncertainty in the softmax output is high (output covering the entire range from 0 to 1). This signifies that the NSM can account for the uncertainty in the prediction. We quantify the uncertainty of the NSM by looking at the entropy of the prediction, defined as $H = -\sum p * log\,(p)$, where $p$ is the probability distribution of the prediction. As shown in Figs. 5(d) and (e), when the NSM makes a correct prediction (classifying image 1 as belonging to class 1), the uncertainty measured in terms of the entropy remains 0. However, in the case of wrong predictions (classifying rotated image of 1 as belonging to class 2 or 4), the uncertainty associated with the prediction becomes large. This is in contrats to the results obtained from a conventional MLP network where the network cannot account for any uncertainty in the data as shown in Fig. 5. Similar results are highlighted when presenting the NSM with rotated images of digit 2 as shown in Figs. 5(f)-(j).

**Conclusion**

Stochasticity works a powerful mechanism in introducing many computational features of a deep neural network such regularization and Monte Carlo sampling. This work builds upon the inherent weight normalization feature exhibited by a stochastic neural network, specifically the Neural Sampling Machine (NSM). Such normalization acts as a powerful feature in most modern deep neural networks [28], [34], [35], mitigating internal covariate shift and providing an alternative mechanism for divisive normalization in bio-inspired neural networks [36]. Our proposed theoretical NSM model provides several advantages: (a) it is an online alternative for otherwise used batch normalization and dropout techniques, (b) it can mitigate saturation at the boundaries of fixed range weight representations, and (c) it provides robustness against spurious fluctuations affecting the rows of the weight matrix.

We demonstrate that the required stochastic nature of the theoretical NSM model can be realized in emerging stochastic devices. This allows seamless implementation of NSM on a hardware using the compute-in-memory architecture. We demonstrate the capability of our proposed hardware NSM to perform image recognition task on standard MNIST dataset with high accuracy (98.25%) comparable to state-of-the-art deterministic neural network. We also showcase the ability of our hardware NSM to perform probabilistic inferencing and quantify the uncertainty in data. Note that while this work focuses on using FeFET as the analog weight cell and $Ag/HfO_2$ as the stochastic selector, a hardware NSM can also be realized using other emerging devices. For example, one can utilize emerging memory candidates such as PCM and RRAM instead of FeFET as the analog weight cell can. For the stochastic selector, other candidates can be explored including Ovonic Threshold Switch (OTS) [37], Mixed Ionic Electronic Conductor (MIEC) [38], and Insulator Metal Transition (IMT) oxides [39] such as Vanadium Dioxide ($VO_2$) [40], [41] and Niobium Oxide ($NbO_x$) [42], [43].

**Methods**

## Fabrication of Ag/HfO₂ stochastic selector

Ag/TiN/HfO₂/Pt devices are fabricated on 250 nm SiO₂/Si substrates. Bottom electrodes are patterned with e-beam lithography and 15nm/60nm Ti/Pt deposited via e-beam evaporation. A 4nm thick HfO₂ film is deposited using atomic layer deposition of TDMAH and H₂O at 120C, followed directly by 3nm thick TiN deposition with TiCl₄ and N₂ at 120C without breaking vacuum. The 150nm thick Ag top electrode is then patterned and deposited using e-beam evaporation, followed by a blanket TiN isolation etch in CHF₃ and electrical testing.

## Calibration of Onrstein-Uhlenbeck (OU) Process

We calibrate the parameters of Eqn. (5) using the experimentally measured threshold voltage $V_T$ of 18 selector devices such as shown in Fig. 3(d). We use the method of linear regression, which has been established in [44] to recast the Eqn. (5) to

$$y = ax + b + \epsilon \tag{8}$$

where $a$ is the slope, $b$ is the interception term and $\epsilon$ is a white noise term. The solution of Eqn. (5) after discretization using the Euler-Maruyama method is given by

$$V_{T_{t+1}} = V_{T_t} e^{-\theta \Delta t} + \mu(1 - e^{-\theta \Delta t}) \sigma \sqrt{\frac{1 - e^{-2\theta \Delta t}}{2\theta}} \mathcal{N}(0,1) \tag{9}$$

By comparing Eqns. (8) and (9), we have $a = e^{-\theta \Delta t}$, $b = \mu(1 - e^{-\theta \Delta t})$ and $sd(\epsilon) = \sigma \sqrt{\frac{1 - e^{-2\theta \Delta t}}{2\theta}}$. Solving for $a$, $b$ and $sd(\epsilon)$, we obtain the OU parameters $\mu = \frac{b}{1-a}$, $\theta = -\frac{\ln a}{\Delta t}$ and $\sigma = sd(\epsilon)\sqrt{\frac{-2 \ln a}{\Delta t(1-a^2)}}$. We have to compute $a$, $b$ and the variance of the error of the linear regression in order to calibrate the OU parameters $\mu$, $\theta$ and $\sigma$. The least square regression terms are $S_x = \sum_{i=1}^{n} S_{i-1}$, $S_y = \sum_{i=1}^{n} S_i$, $S_{xx} = \sum_{i=1}^{n} S_{i-1}^2$, $S_{xy} = \sum_{i=1}^{n} S_{i-1} S_i$ and $S_{yy} = \sum_{i=1}^{n} S_i^2$ where $S$ represents a sample drawn from the experimental data. Upon further simplification, we end up with computing the following equations

$$a = \frac{nS_{xy} - S_x S_y}{nS_{xx} - S_x^2} \tag{10}$$

$$b = \frac{S_y - aS_x}{n} \tag{11}$$

$$sd(\epsilon) = \sqrt{\frac{nS_{yy} - S_y^2 - a(nS_{xy} - S_x S_y)}{n(n-2)}} \tag{12}$$

The parameter $\sigma$ is computer as the ratio of $\frac{sd(\epsilon)}{\sqrt{\Delta t}}$, where $\Delta t$ is the sampling step for the experimental data or the time step of the Euler-Maruyama method.

## Training Process of NSM

The multilayer perceptron (MLP) network described in Fig. 4(a) was trained with the backpropagation algorithm [45], the Cross-entropy as loss function and an adapted version of Adam optimizer with a learning rate of 0.0003 and betas (0.9, 0.999). We adapted the Adam optimizer to accommodate for the updates of the conductance in the FeFet model (see paragraph: Building Blocks for Stochastic Synapse: FeFET-based Analog Weight Cell). The training and testing batch sizes were both set to 100. We trained the network for 200 epochs and at each epoch we used the full 60,000 samples training MNIST set. The learning rate was linearly decreased after 100 epochs with a rate of $0.0003 \times min\left\{2 - \frac{x}{100}, 1\right\}$, where $x$ is the number of a specific epoch. Every two epochs we measured the accuracy of the network using the full 10,000 samples testing MNIST set over an ensemble of 100 samples of the forward pass of the neural network. The accuracy was measured as the ratio of successfully classified digits to the total number of samples within the test MNIST set (10,000). All the experiments ran on a Nvidia GPU Titan X with 12GB of physical memory and a host machine equipped with a Intel i9 with 64 GB physical memory running Arch Linux. The source code is written in Python (Pytorch, Numpy, Sklearn) and it will [be freely available on-line upon acceptance for publication].

## Data availability

The data that support the findings of this study are available from the corresponding author upon request.

## Code availability

The simulation codes used for this study are available from the corresponding author upon request.

**Acknowledgements**

We are grateful to M. Trentzsch, S. Dunkel, S. Beyer, and W. Taylor at Globalfoundries Dresden, Germany for providing 28nm HKMG FeFET test devices. This project was supported by the National Science Foundation (NSF), and the Nanoelectronics Research Corporation (NERC), a subsidiary of the Semiconductor Research Corporation (SRC), through Extremely Energy Efficient Collective Electronics (EXCEL).

**Author Contribution**

S. Dutta, G.D., E.N. and S. Datta developed the main idea. S. Dutta and A.K. performed all the measurements. B.G. helped with fabrication of the selector devices. G.D. and E.N. performed the simulations for NSM. All authors discussed the results, agreed to the conclusions of the paper and contributed to the writing of the manuscript.

**Figures**

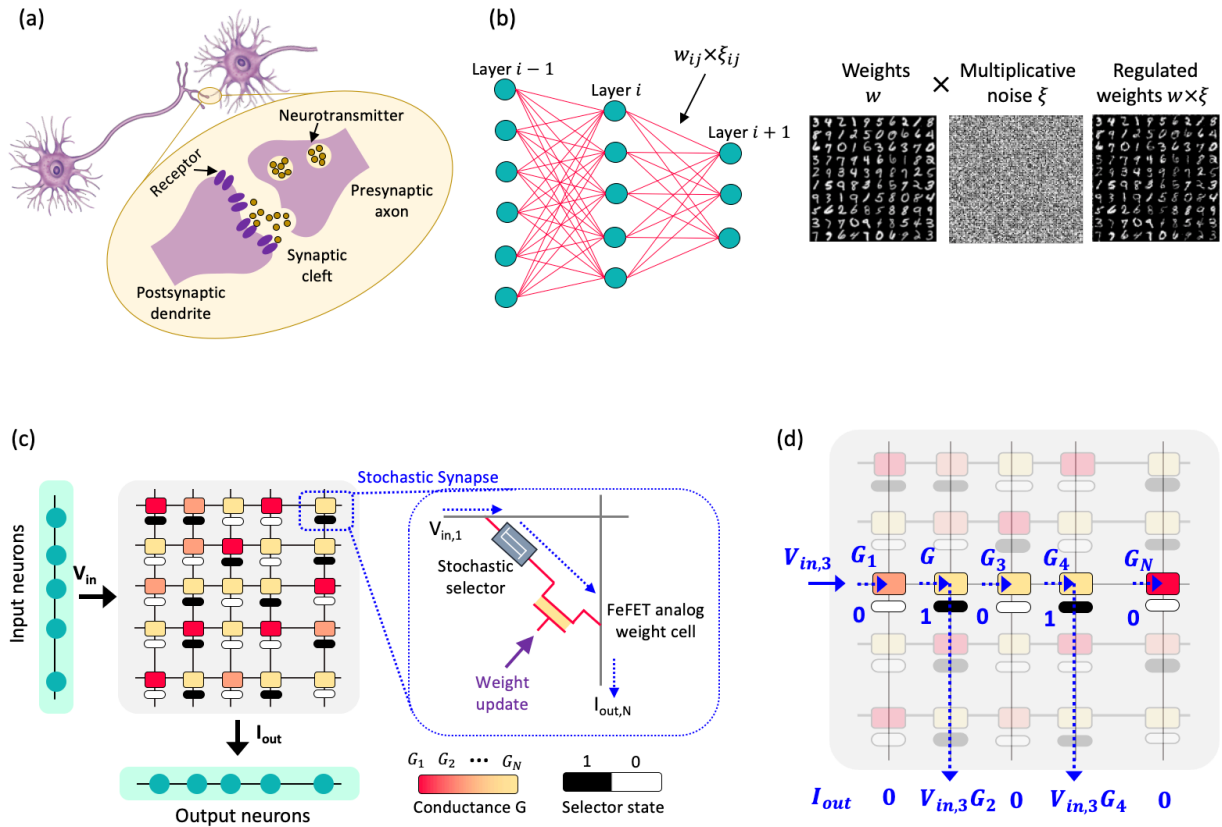

**Figure 1.** (a) Synaptic stochasticity occurring at the molecular level in biological neural networks. The presynaptic neuronal spike causes the release of neurotransmitters at the synaptic release site with a probability around 0.1. (b) Schematic of a Neural Sampling Machine (NSM) incorporating a Bernoulli or *"blank-out"* multiplicative noise in the synapse. This acts as a continuous DropConnect mask on the synaptic weights such that a subset of the weights is continuously forced to be zero. (c) Illustration of an NSM implemented in a hardware using crossbar array architecture implementing compute-in-memory. The analog weight cell implemented using eNVMs are placed at each cross-point and are augmented with a stochastic selector element. This allows selectively sampling or reading the synaptic weights $G_{ij}$ with some degree of uncertainty, based on random binary variables $\xi_{ij}$ generated for each of the synapse. (d) Illustration of a scenario where an input voltage $V_{in,3}$ is applied to a row of the synaptic array with conductance states $\boldsymbol{G} = \{G_1, G_2, G_3, G_4, \ldots, G_N\}$. Depending on the state of the selectors in the cross-points, an output weighted sum current $\boldsymbol{I_{out}} = \{0, G_2 V_{in,3}, 0, G_4 V_{in,3}, \ldots, 0\}$ is generated which is exactly same as multiplying the weight sum of $w_{ij} z_j$ with a multiplicative noise $\xi_{ij}$

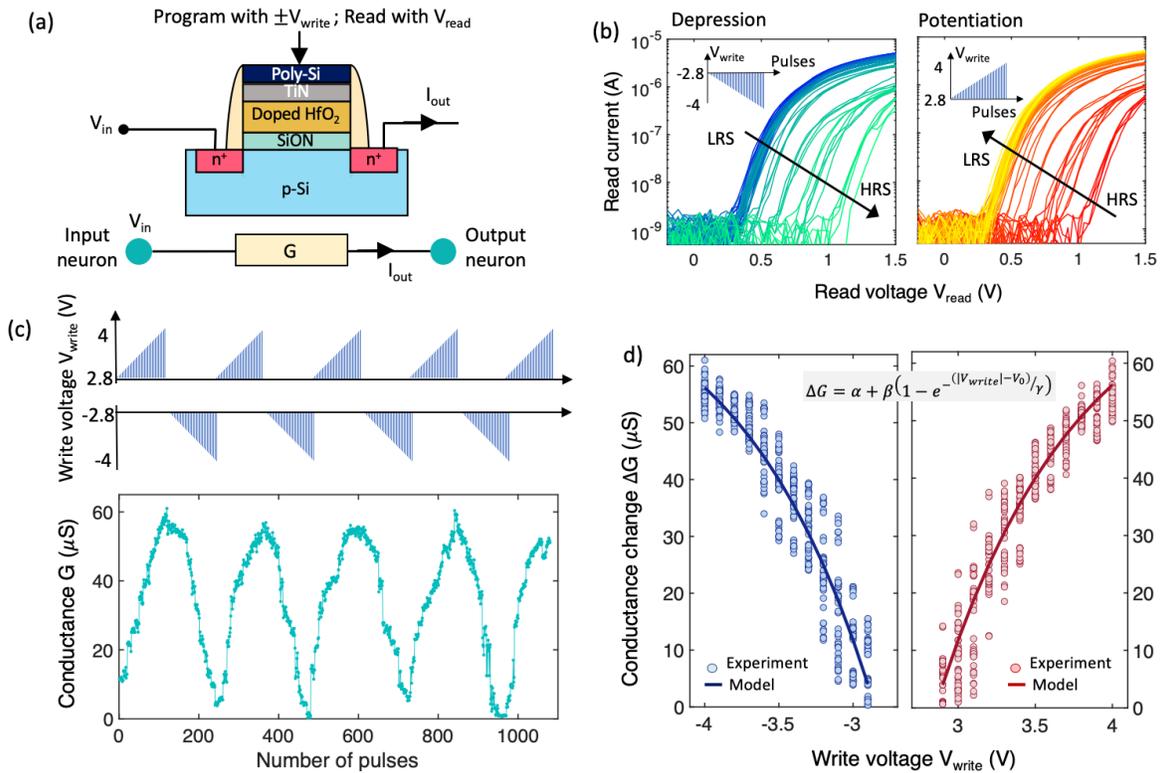

**Figure 2.** (a) Schematic of a stand-alone FeFET-based analog synapse. The channel conductance can be modulated by applying write pulses $\pm V_{write}$ to the gate of the FeFET while reading out the conductance state is achieved by applying a small read voltage $V_{read}$ to the gate terminal. (b) Experimentally measured conductance modulation in a 500 nm x 500 nm high-K metal gate FeFET fabricated at 28nm technology node. An amplitude modulation scheme is used where positive and negative write voltage pulses $V_{write}$ of increasing amplitude from 2.8V to 4V and pulse widths of 1$\mu$s are applied to modulate the conductance of the FeFET. (c) Measured continuous change in the conductance state of the FeFET upon applying multiple potentiation and depression pulses of varying amplitude. (d) The FeFET-based analog weight cell is modeled in the NSM by fitting the conductance update scheme for both potentiation and depression with the closed-form expression as shown in the figure.

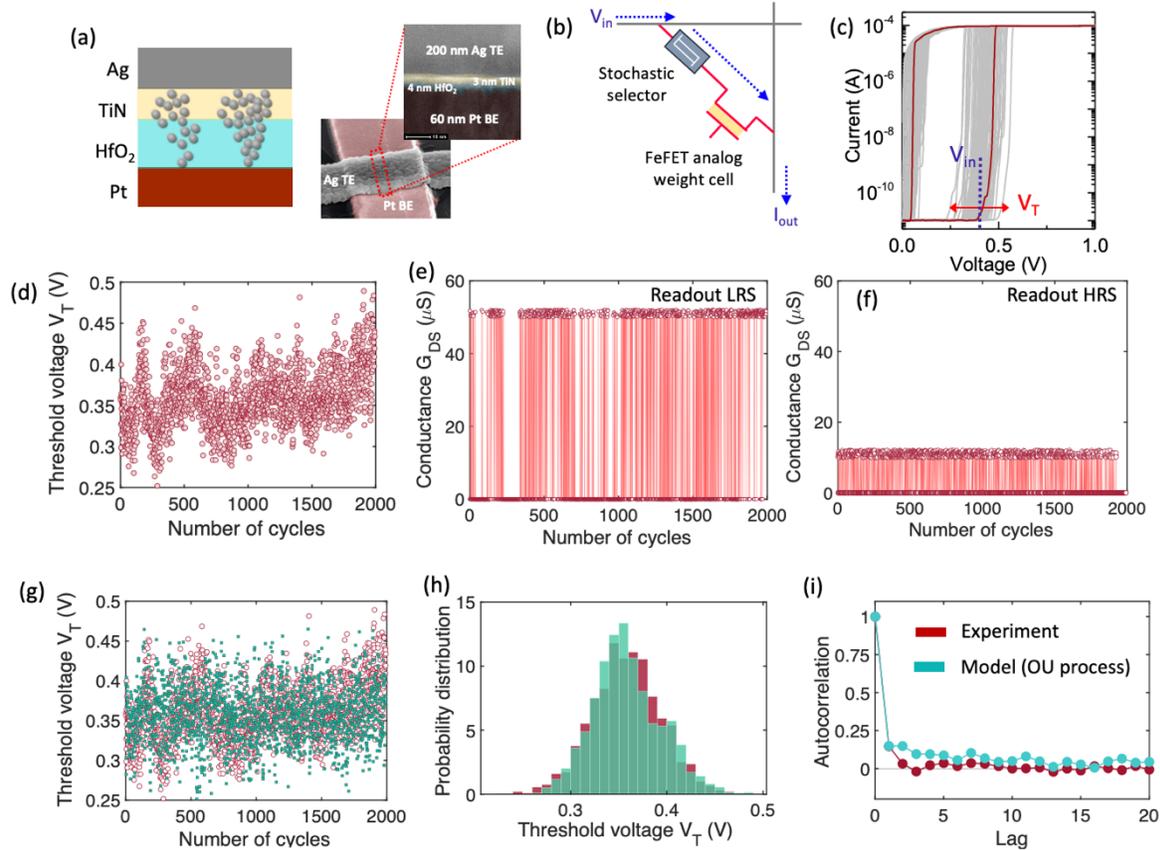

**Figure 3.** (a) Schematic and TEM of a fabricated stack of [Ag/TiN/HfO$_2$/Pt] with 3nm TiN and 4nm HfO$_2$. (b) A stochastic synapse is realized by augmenting this stochastic selector in series with the FeFET-based analog weight cell. (c) Measured current-voltage characteristics showing abrupt electronic transition from insulating state to metallic state due to the formation of a continuous filament of Ag+ atoms bridge the top and bottom electrodes. A wide window of variation in the threshold voltage V$_T$ that triggers the spontaneous formation of the Ag$^+$ filament is observed. The stochasticity can be exploited by applying the input voltage V$_{in}$ within the variation window of the V$_T$. (d) Measured threshold voltage V$_T$ over multiple cycles. (e, f) Stochastically reading an LRS and an HRS of the FeFET through the stochastic selector. (g-i) The stochasticity switching of the selector device is modeled using an Onrstein-Uhlenbeck (OU) Process. The model shows excellent agreement with the experimental data.

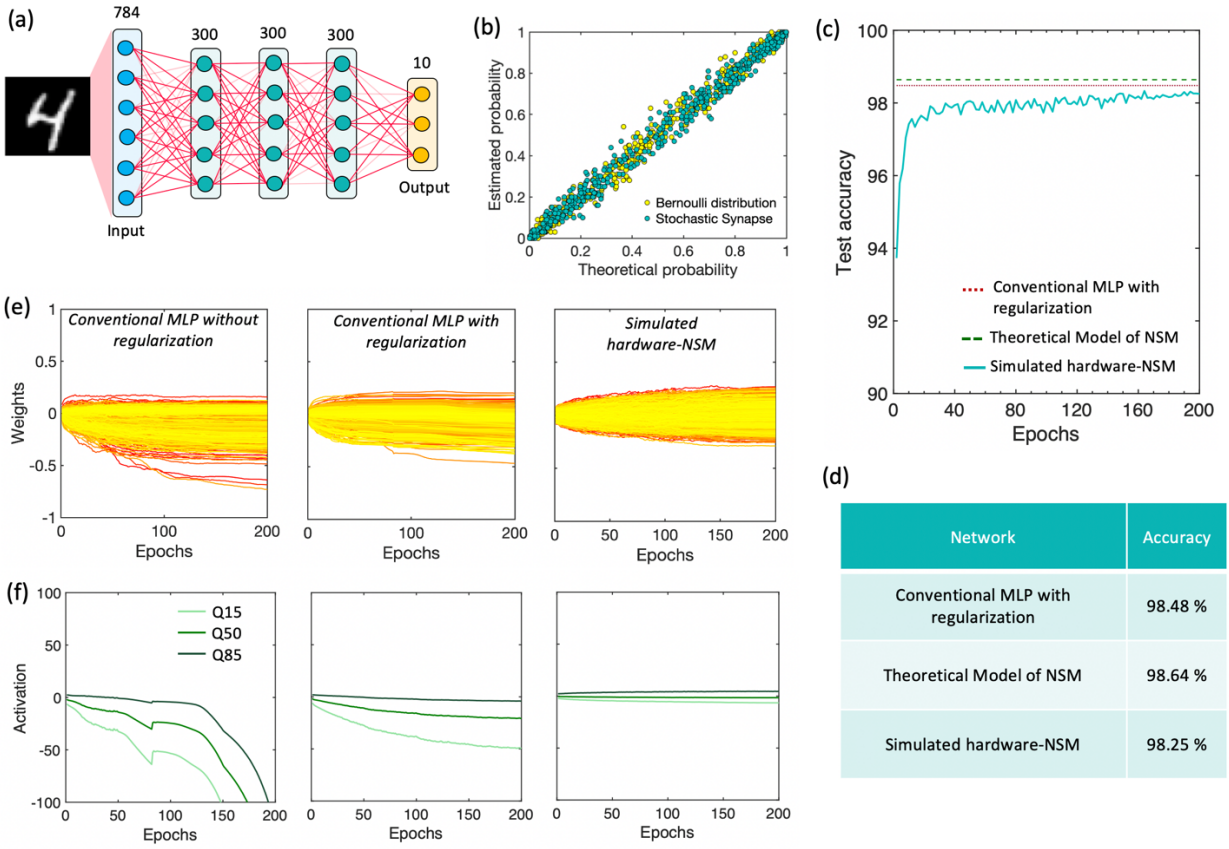

**Figure 4.** (a) Network architecture of the NSM consisting of an input layer, three hidden fully connected layers and an output layer. (b) Exact match witnessed between the measured switching probability of the stochastic selector device and theoretically predicted probability for a Bernoulli distribution, highlighting that our stochastic selector device can inject Bernoulli multiplicative noise. (c) Evolution of the test accuracy for the simulated hardware-NSM using the FeFET-based analog weight cell and the stochastic selector as a function of the epochs. (d) Comparison of the performance of the simulated hardware-NSM with a deterministic feedforward multilayer perceptron (MLP) and the theoretical NSM model with full precession synaptic weights and a Bernoulli multiplicative noise for the stochastic synapses. (e) Evolution of the weights of the third layer during learning for three different networks- an MLP without any regularization, an MLP with additional regularization added and the simulated hardware-NSM. (f) Evolution of the 15th, 50th and 85th percentiles of the input distributions to the last hidden layer during training for all the three networks. Overall, NSM exhibits a tighter distribution of the weights and activation concentrated around its mean, highlighting the inherent self-normalizing feature.

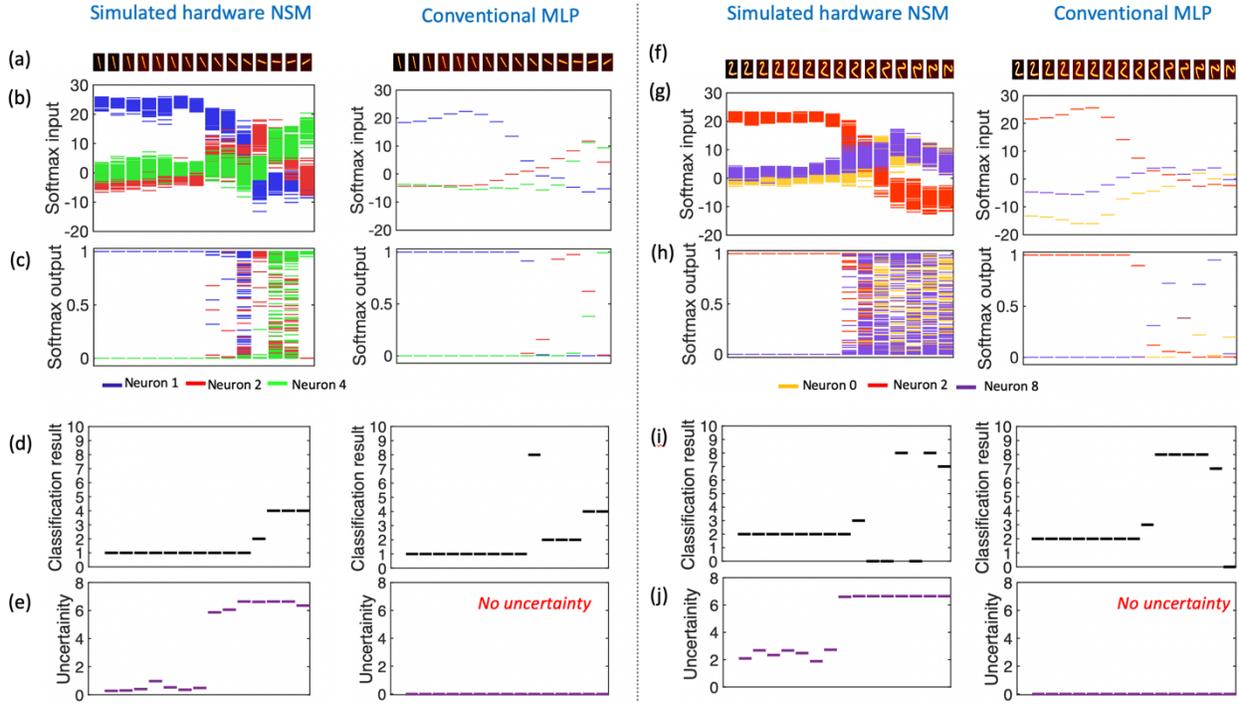

**Figure 5.** Bayesian inferencing and uncertainty in data comparison between simulated hardware-NSM and a conventional MLP network. (a, f) Continuously rotated images of the digits 1 and 2 from the MNIST dataset, used for performing Bayesian inferencing. We perform 100 stochastic forward passes during the inference mode for each rotated image of digits 1 and 2 and record the distribution of the (b, g) softmax input and (c, h) softmax output for few representative output neurons. (d, i) Classification result produced by the NSM for each rotated image. (e, j) The uncertainty of the NSM associated with the prediction, calculated in terms of the entropy $H = -\sum p * log\,(p)$, where $p$ is the probability distribution of the prediction. When the NSM makes a correct prediction (classifying image 1 and 2 as belonging to class 1 and 2, respectively), the uncertainty measured in terms of the entropy remains 0. However, in the case of wrong predictions, the uncertainty associated with the prediction becomes large.